\documentclass[14pt]{article}
\usepackage{subcaption,amsmath,tabularx,cite}
\usepackage[a4paper]{geometry}
\usepackage{fancyhdr}
\usepackage{ifthen}
\usepackage{graphicx}

\usepackage{indentfirst}
\usepackage[colorlinks=true,urlcolor=black,linkcolor=black,filecolor=black,citecolor=black]{hyperref}

\usepackage[titletoc]{appendix}
\usepackage{authblk}
\title{Prospects for establishing limits on the SMEFT operators from the production processes of three and four top quarks in hadron collisions}
\author[1]{A.Aleshko}
\author[1]{E.Boos}
\author[1]{V.Bunichev}
\author[1]{L.Dudko}
\affil[1]{Skobeltsyn Institute of Nuclear Physics, Lomonosov Moscow State University, 119991 Moscow, Russia}
\begin{document}

\maketitle




\begin{abstract}

Numerical simulations of processes of three and four top quark hadroproduction are carried out in the SMEFT model framework. The simulated data are used to derive expected theoretical constraints on Wilson coefficients of relevant SMEFT operators of
dimension six. Obtained limits for both cases are discussed and compared in terms of processes' sensitivity to possible BSM contribution. Results show that operator $O_{tt}^1$ is better constrained by the process of four top quark production, whereas other four operators $O_{QQ}^1$, $O_{Qt}^1$, $O_{Qt}^8$ and $O_{QQ}^8$, are similarly constrained in three and four top quark production processes. In all cases, the expected limits taken from the simultaneous analysis of the production of three and four top quarks are strengthened. Analytical expressions for the partial amplitudes of the processes $tt\to tt$ and $t\bar{t}\to t\bar{t}$ caused by the operators $O^1_{tt}$, $O^1_{QQ}$, $O^1_{Qt}$, $O^8_{Qt}$, $O^8_{QQ}$ were obtained for the first time. Based on the expressions of the obtained partial amplitudes, graphs of the perturbative unitarity boundary for the listed operators were drawn. The question of how kinematic cuts motivated by partial unitarity affect the resulting constraints on the Willson coefficients is addressed. It is shown that in all cases the limits are getting somewhat worse if such cuts are applied. 

\end{abstract}

\section{Introduction}

Currently, no experimental evidence of physics beyond the Standard Model (BSM) has been observed. In the pursuit of the New Physics, researchers are inclined to try ``indirect'' approaches, in which one seeks BSM manifestations in the interactions of already known Standard Model particles. The conventional assumption here is that New Physics is on a scale beyond our direct reach at the moment, but should still manifest itself on lower scales in the form of modified SM interactions, which can be measured and analyzed. A convenient framework for such kind of analyzes is the Standard Model Effective Field Theory (SMEFT)~\cite{Weinberg:1979sa,Buchmuller:1985jz,Grzadkowski:2010es,Degrande:2012wf,Alonso:2013hga,Aguilar-Saavedra:2018ksv} (see also reviews~\cite{Gripaios:2015qya,Boos:2022cys} and references therein). SMEFT approach is to parametrize BSM effects in a model-independent way in terms of higher dimension gauge-invariant operators. If only operators of dimension six are preserved, then the SMEFT Lagrangian reads as follows:

\begin{equation}
    L = L_{SM} + \sum^{}_{}{\frac{c_i}{\Lambda^2}O^{d=6}_i}, 
\label{eft_lagr}
\end{equation}

\noindent where $L_{SM}$ is the SM Lagrangian, $\Lambda$ - hypothetical scale of the BSM physics, $O^{d=6}_i$ - local composite SMEFT operators of dimension six, $c_i$ - dimensionless Wilson coefficients. We consider here only terms of dimension six since in the processes under study they are leading in the expansion on inverse scale $\Lambda^{-1}$. In the SMEFT framework any observable, in particular the cross-section, can be parametrized in the following form:
\begin{equation}
    \sigma = \sigma_{SM} + \sum^{}_{k}{\frac{c_i}{\Lambda^2}\sigma^{(1)}_k} + \sum^{}_{j <= k}{\frac{c_i c_k}{\Lambda^4}\sigma^{(2)}_{k,j}},
\label{sigma_eq}
\end{equation}
where \( \sigma_{SM} \) is the SM value, $\sigma^{(1)}$ and $\sigma^{(2)}$ - coefficients, representing linear and quadratic (in terms of EFT coupling) contributions of the SMEFT operators. Given the measurement of $\sigma$ and having calculated values of $\sigma_{SM}$, $\sigma^{(1)}$ and $\sigma^{(2)}$, one can estimate constrains on Wilson coefficients $c_i$. These constraints can be used to calculate limits on physical parameters in various SM extensions.

Search for four top quark production was performed and upper limits on its cross-section have been established~\cite{CMS:2019jsc, CMS:2019rvj,ATLAS:2018kxv} in proton-proton collisions at $\sqrt{s}=13$ TeV. The process was observed recently with measured cross-section $17.7^{+6.0}_{-5.4}$ fb~\cite{CMS:2023ftu} by CMS and $22.5^{+6.6}_{-5.5}$ fb~\cite{ATLAS:2023ajo} by ATLAS experiments. 
The SM NLO QCD cross section at 14 TeV has been calculated~\cite{Bevilacqua:2012em} for various choices of the factorization and renormalization scales.
In addition, the computation of NLO QCD and EW corrections in SM have been carried out in~\cite{Frederix:2017wme} at 13 and 100 TeV energies.
The SM cross section at next-to-leading logarithmic accuracy including threshold non-logarithmic corrections  (so-called NLL$^\prime$) is now available for 13 and 13.6 TeV~\cite{vanBeekveld:2022hty}.
Complete tree level QCD and Electroweak analysis including contributions of all five relevant SMEFT operators of dimension six was presented and expected individual theoretical limits were given at various energies~\cite{Aoude:2022deh}. CMS and ATLAS collaborations also presented experimental limits on corresponding dimension six operators~\cite{CMS:2019jsc,ATLAS:2023ajo}.
One should stress that all the computations of the four top quark production cross sections at corresponding energies are consistent with each other and with the mentioned experimental measurements within claimed uncertainties.

One can also consider processes with 3, 5, etc. top quarks in a final state when searching for SMEFT manifestation beside the four top production. The four top quark production is mostly due to the QCD contribution. The three top quark production, on the other hand, always contain EW vertices, which results in a lower cross-section, but also a lower background. This suggests that, despite the lower cross-section, the triple top production may still be a fairly good target for BSM studies. In the current work, we consider the process of three top quark production and compare its sensitivity to the contributions of the SMEFT operators with that obtained in the relatively well-studied four top quark production process, and also investigate how the sensitivity improves when both processes are taken into account simultaneously.

Three top quark production has LO SM cross-section of 1.9 fb~\cite{Boos:2021yat, Barger:2010uw} and has not been experimentally observed yet. Nevertheless, one can estimate possible constraints on relevant Wilson coefficients following from three top quark production by comparing SM and SMEFT theoretical cross sections. We use the SM cross section as an effective ``measurement'' and compare it with the SMEFT cross-section as given in Eq.~\ref{sigma_eq}. In order to compare expected sensitivities obtained from the three and four top quark production processes, the same analysis procedure is applied to both processes to determine the limits. 
The expected limits obtained are also compared with currently known experimental limits only for the production of four top quarks~\cite{CMS:2019jsc,ATLAS:2023ajo}.

One potentially serious problem arising in the SMEFT approach is that the contributions of EFT operators grow too fast with energy. An important consequence is that one should be very careful with the potential violation of unitarity. In the current work, we investigate this problem by studying the effect of the partial unitarity requirement on the limits of the extracted Wilson coefficients. We also obtain the corresponding kinematic cutoff, which can be used in simulations to ensure partial unitarity.

Thus, the work has two main goals: to estimate the theoretically expected limits on the Wilson coefficients obtained in the process of the production of three and four top quarks and to study the impact of the unitary requirement. The paper is organized as follows. Section~\ref{Sec:cs} contains computation details, cross sections and theory uncertainties.
The partial unitarity limits are discussed in Section~\ref{unitarity}. In Section~\ref{sec:limits} the methodology of obtaining limits has been discussed. The final results and comparisons are presented in the Summary.

\section{Cross sections and theoretical uncertainties}
\label{Sec:cs}

\subsection{Simulation details}

In the present work, most of the calculations are conducted using the MadGraph5\_aMC\@NLO package~\cite{madgraph}. 
Some of the results were also cross-checked with the CompHEP package~\cite{comphep, Pukhov:1999gg}. For modeling with SMEFT operator contributions, we use the SMEFTatNLO model~\cite{smeft_at_nlo}. Simulations at LO and NLO are done using NNPDF31\_lo\_as\_0118 and NNPDF31\_nlo\_as\_0118\_luxqed~\cite{nnpdf} PDF sets respectively. The value of $\alpha_{s}$ is taken as follows from the PDF set. The mass of the top quark is set to 172.5 GeV. To obtain factorization/renormalization scale uncertainties, we perform calculations at various characteristic scale points as implemented in MadGraph. All other settings are taken as default in Madgraph unless stated otherwise.

\subsection{Computations in Standard Model}
\label{Sec:cs:sm}

\begin{figure}[t!]
\centering
\includegraphics[width=0.24\textwidth,clip]{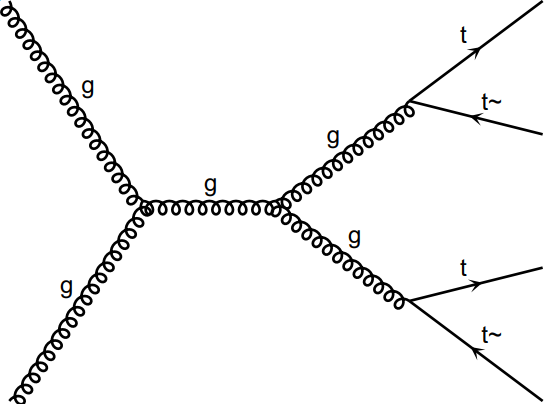}
\includegraphics[width=0.24\textwidth,clip]{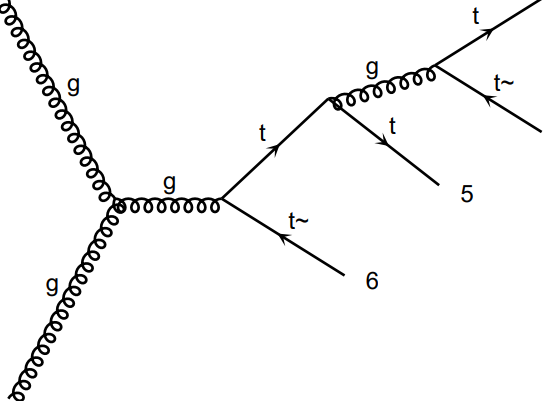}
\includegraphics[width=0.24\textwidth,clip]{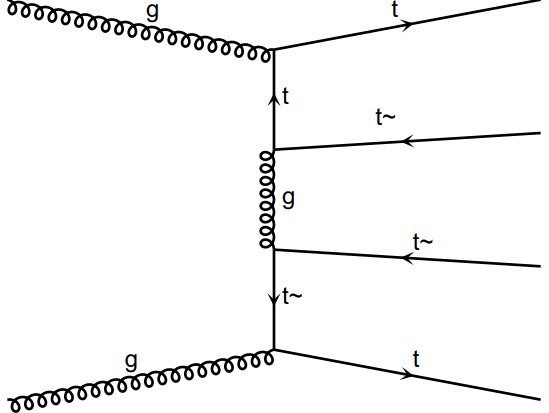}
\includegraphics[width=0.24\textwidth,clip]{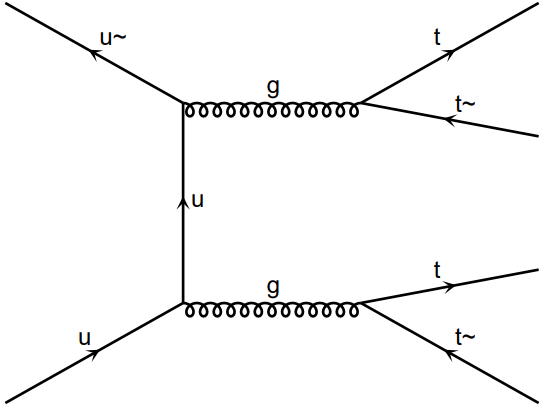}
\caption{Representative Feynman diagrams of four top quarks LO production in the Standard Model.}
\label{4topsm}
\end{figure}

\begin{figure}[t!]
\centering
\includegraphics[width=0.24\textwidth,clip]{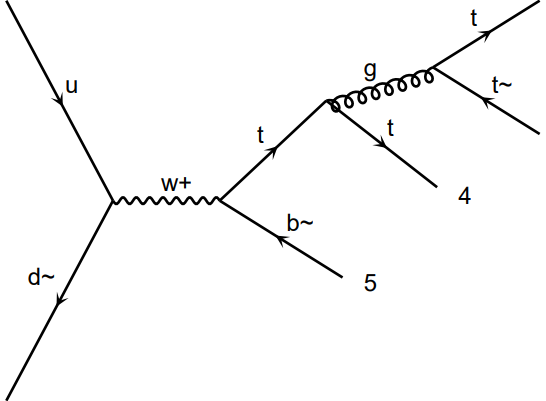}
\includegraphics[width=0.24\textwidth,clip]{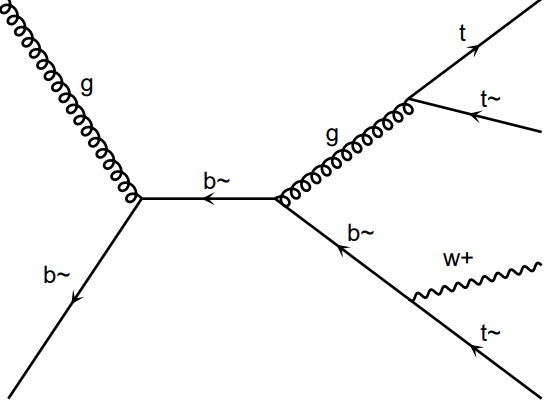}
\includegraphics[width=0.24\textwidth,clip]{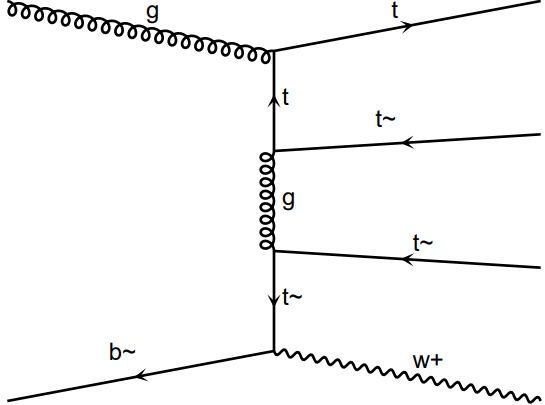}
\caption{Representative Feynman diagrams of three top quarks LO production in the Standard Model.}
\label{3topsm}
\end{figure}

Representative LO diagrams for four top SM production are depicted in figure~\ref{4topsm}.
In our study, we use NLO corrections to the cross section of four top quark production as calculated in SM~\cite{Bevilacqua:2012em, Frederix:2017wme}. The numerical value of the cross-section strongly depends on the factorization/renormalization scale $\mu_{F/R}$ and the PDF set chosen. In table~\ref{t1} SM cross sections at $\sqrt{s} = 13$ TeV are presented. We explore different choices of factorization/renormalization scales, which are often used for this process: $H_{t}/2$, $2m_{top}$, and $m_{top}$. $H_{t}/2$ is defined as half of a scalar sum of transverse momenta of all final state particles, while the $m_{top}$ corresponds to the fixed scale equal to the mass of the top quark. With the current PDF set, the latter choice of scale provides a smaller K-factor, hence we will use it for all subsequent computations. 

\begin{table}[t]
\begin{center}
\begin{tabular}{ c|c|c|c } 
\hline 
{Order} & \multicolumn{3}{c}{LO} \\ \hline 
 Scale, $\mu_{F/R}$ & cross-section, $\sigma$ pb & $\delta_{\rm scale}$ & $\delta_{PDF}$ \\ 
 \cline{2-4}
 \(m_{top} \)     & 9.03 & +73\% -39\% & $\pm$7.3\% \\
 \(2m_{top} \)     & 5.47 & +65.1\% -36.9\%  & $\pm$ 6.81\% \\
 \(H_{t}/2 \)     & 3.93 & +60\% -35\% & $\pm$ 6.4\% \\
 \hline
 \hline
 {Order} & \multicolumn{3}{c}{QCD NLO} \\
 \hline
 Scale, $\mu_{F/R}$ & cross-section, $\sigma$ pb & $\delta_{scale}$ & $\delta_{PDF}$ \\ 
  \cline{2-4}
 \(m_{top} \)     & 13.2 & +9.5\% -20.1\%  & $\pm$ 2.7\% \\
 \(2m_{top} \)     & 11.2 & +26.7\% -25.0\%  & $\pm$ 2.6\% \\
 \(H_{t}/2 \)     & 8.4  & +29.6\% -25.1\% & $\pm$ 2.5\% \\
 \hline
\end{tabular}			
\end{center} 			
\caption{Cross-sections of the four top quarks hadroproduction with corresponding scale/PDF uncertainties.}
\label{t1}
\end{table}

The three top quark production is also been studied in some works~\cite{ Barger:2010uw, Malekhosseini:2018fgp, Khanpour:2019qnw, Boos:2021yat}. Representative diagrams for three top LO SM production are shown in figure~\ref{3topsm}. A full set of diagrams can be found in \cite{Boos:2021yat}. The three top production is also strongly dependent on the choice of QCD factorization scale, the main consequence of which is a relatively large scale uncertainty. To correctly compare the two processes we also want to use NLO corrections to the cross section of three top quark production, similar to the four top quark production. The calculation of triple top production with NLO corrections is tricky due to the presence of resonant diagrams in the real part of the correction. These are diagrams of type $pp \rightarrow t\bar t t\bar t$ with decay $t \rightarrow Wb$. LO cross-section for the production of four top quarks is an order of magnitude greater than one for triple top production, which therefore leads to spoilage of convergence of the perturbative series. From a computational point of view, this means that the simulation will produce very unstable results (or will not converge at all). To obtain precise results, the issue should be considered in the same way as $tWb$ production processes, which is beyond the scope of the current work. Therefore, for the current exploratory goal of this paper, we simply use the general Diagram Reduction procedure as implemented in the MadSTR plugin~\cite{madstr} for Madgraph. Table~\ref{t2} contains results for three and four top quarks SM production at c.o.m. energies 13 and 14 TeV.

\begin{table}[t]
    \resizebox{0.9\textwidth}{!}{%
    \begin{tabular}{l|c|c|c}
    \multicolumn{4}{c}{4 top production, LO} \\
    \hline
       C.o.m. energy, TeV & SM cros.-sect. $\sigma_{SM}$, fb & scale uncert., \% & PDF uncert., \% \\  \hline
       13 & 9.02 & 73.4 &	7.32 \\
       14 & 12.0 & 72.6 &	7.28 \\
       
    \hline
    \multicolumn{4}{c}{4 top production, NLO} \\
    \hline
       13 & 13.2 & 20.1 &	2.5 \\
       14 & 17.8 & 20.3 &	2.5 \\

    \hline
    \hline
    \multicolumn{4}{c}{} \\
    \multicolumn{4}{c}{3 top production, LO} \\
    \hline
       C.o.m. energy, TeV & SM cros.-sect. $\sigma_{SM}$, fb & scale uncert., \% & PDF uncert., \% \\  \hline
       13 & 1.16 & 30.1 &	7.1 \\
       14 & 1.5 & 29.2 &	6.64 \\
       
    \hline
    \multicolumn{4}{c}{3 top production, NLO} \\
    \hline
       13 & 1.78 & 20.0 &	3.3 \\
       14 & 2.3 & 19.4 &	3.1 \\

    \end{tabular}
    }
    \caption{SM cross-sections for processes $pp \rightarrow t \bar{t} t \bar{t}$ and $pp \rightarrow t \bar{t} \bar{t} (t \bar{t} t)$}
    \label{t2}
\end{table}

\subsection{SMEFT computations}

The introduction of SMEFT operators adds new possible interaction vertices, which modifies SM cross-sections. There are only 5 dimension-six four-fermion SMEFT operators, which contribute to the four top quark production process:

\begin{align} \label{eft_opetators}
    O_{tt}^{1} &= (\bar{t}_R\gamma^{\mu}t_R)(\bar{t}_R\gamma_{\mu}t_R), \\ \nonumber
    O_{QQ}^{1} &= (\bar{Q}_L\gamma^{\mu}Q_L)(\bar{Q}_L\gamma_{\mu}Q_L), \\ \nonumber
    O_{Qt}^{1} &= (\bar{Q}_L\gamma^{\mu}Q_L)(\bar{t}_R\gamma_{\mu}t_R), \\ \nonumber
    O_{Qt}^{8} &= (\bar{Q}_L\gamma^{\mu}T^{A}Q_L)(\bar{t}_R\gamma_{\mu}T^{A}t_R), \\ \nonumber
    O_{QQ}^{8} &= (\bar{Q}_L\gamma^{\mu}T^{A}Q_L)(\bar{Q}_L\gamma_{\mu}T^{A}Q_L), \\ \nonumber
\end{align}

\begin{figure}[t]

\begin{subfigure}[tp]{0.48\textwidth}
    \centering
    \includegraphics[width=0.48\textwidth,clip]{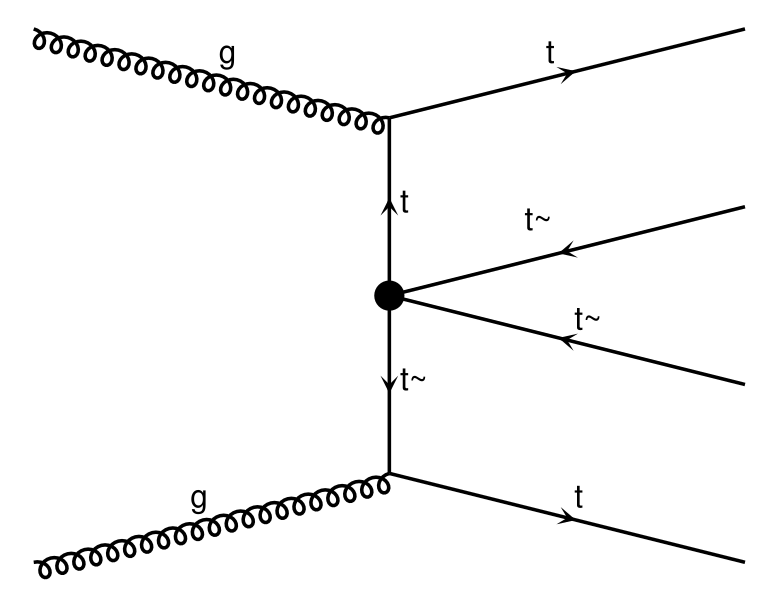}
    \includegraphics[width=0.48\textwidth,clip]{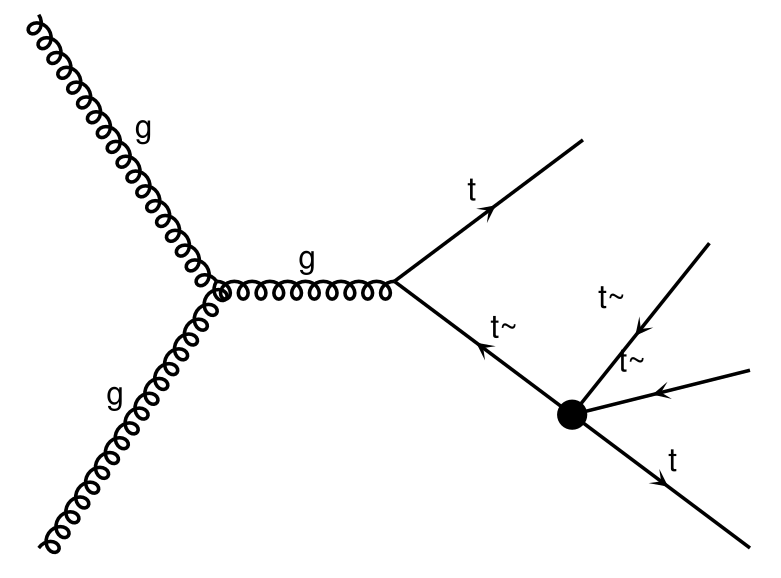}
    \caption{ }
\end{subfigure}
\begin{subfigure}[tp]{0.48\textwidth}
    \centering
    \includegraphics[width=0.48\textwidth,clip]{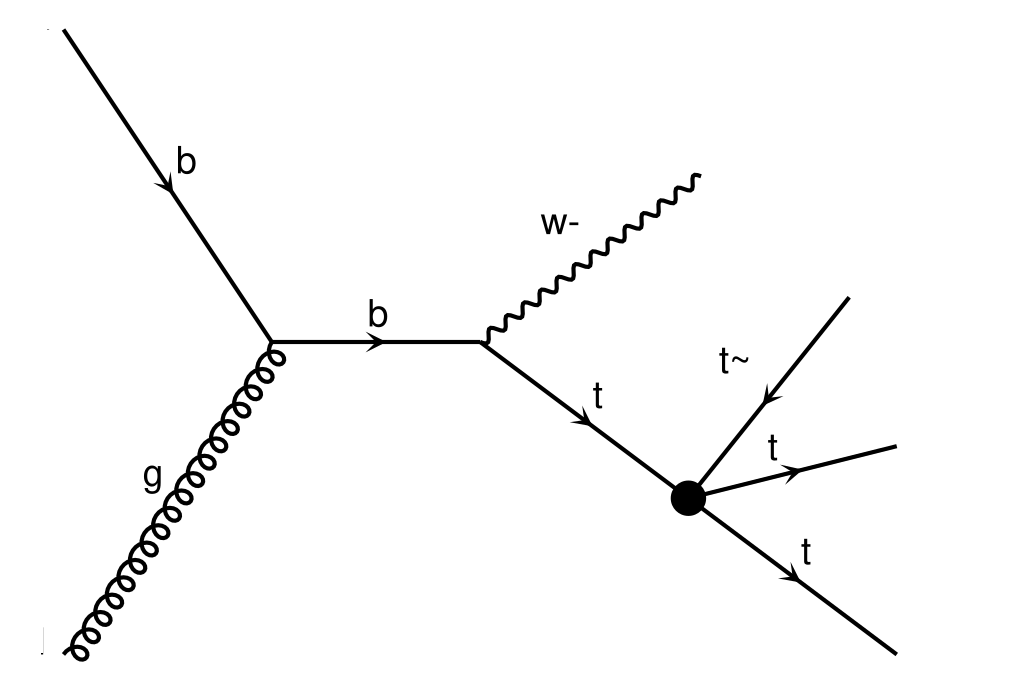}
    \includegraphics[width=0.48\textwidth,clip]{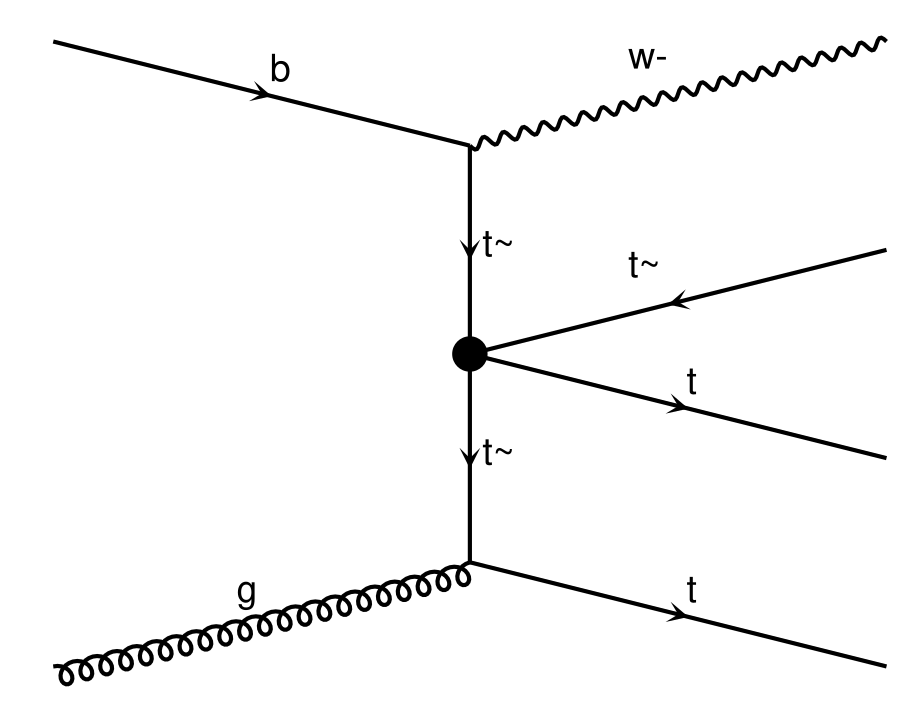}
    \caption{ }
\end{subfigure}

\caption{Examples of Feynman diagrams with SMEFT vertices for the four (a) and the three (b) top quark hadroproduction.}
\label{top_eft}
\end{figure}

Since much of this analysis is based on comparisons between the four and the three top quark production processes, we will consider only this set of five operators in what follows. Some representative SMEFT Feynman diagrams with operators from list \ref{eft_opetators} can be seen in Fig. \ref{top_eft}

The SMEFT model introduces several additional parameters, namely the New Physics scale $\Lambda$ and Wilson coefficients $c_i$ for each SMEFT operator. The scale $\Lambda$ is conventionally set to 1 TeV, while $c_i$ is our "parameters of interest", which we want to set bounds on. 

The main idea of this work is to use the theoretical value of the cross-section as an approximation for the experimental value in Eq.~\ref{sigma_eq} to obtain theoretical constraints on the Wilson coefficients of operators listed in Eq.~\ref{eft_opetators}.

To extract values for parameterization coefficients $\sigma^{(1)}$ and $\sigma^{(2)}$ a series of simulations using SMEFT model was conducted. The latest versions of Madgraph allow for direct computations of squared and interference SMEFT terms, however, it is not advised, as computation of small interference parts can potentially be unstable. Therefore following procedure of coefficient extraction is utilized. Our approach is to calculate the cross-section with two opposite values of $c_i$ (i.e. $c_i$ and $-c_i$). In principle, one can take an arbitrary value of $c_i$, since it only affects coefficients in the set of linear equations generated by \ref{sigma_eq}, which when solved for $\sigma^{(1)}$ and $\sigma^{(2)}$ provide the same values, regardless of the choice. One should note, however, that due to the choice of $\Lambda$ the value of $c_i$ is required to stay within the $[-4\pi,4\pi]$ range to ensure the stability of the perturbation series. We tested several choices of $c_i$ and settled for the simple $c_i = \pm 1$ for the sake of clarity. Hence, to obtain coefficients $\sigma^{(1)}$ and $\sigma^{(2)}$ we calculate cross-section twice with $c_i = \pm 1$ and solve linear equation set obtained from \ref{sigma_eq}. 

The described procedure is conducted for all of the SMEFT operators listed in~\ref{eft_opetators} as well as for both three and four top production processes. Thus obtained values for $\sigma^{(1)}$ and $\sigma^{(2)}$ are used in section~\ref{sec:limits} to obtain limits for Wilson coefficients $c_i$. However, before moving to the calculation of restriction on Wilson coefficients, the important point to discuss is the problem of potential violation of perturbative unitarity.


\section{Optical theorem and perturbative unitarity}
\label{unitarity}
Effective operators lead to an increase in cross sections with an increase in energy, which violates unitarity. For our calculations to be self-consistent, we must check that we do not consider kinematic regions where perturbative unitarity is violated.
To estimate the admissible range of parameters, we apply the optical theorem, which follows from the unitarity of the S matrix. The optical theorem states that the imaginary part of the forward scattering amplitude is proportional to the total cross section of the process:

\begin{align}\label{unit_formula1}
	\sigma = \frac{1}{s}{\rm Im}\left(A(\theta=0)\right)=\frac{16\pi}{s}\sum\limits^{\infty}_{l=0}(2l+1)|a_l|^2,
\end{align}
where $a_l$ – is the amplitude of the partial wave. Therefore, Im$a_l=|a_l|^2$ and 

\begin{align}\label{unit_formula2}
	|\rm Re(a_l)|^2 + \left[\rm Im(a_l)-\frac{1}{2} \right]^2 = \frac{1}{4}.
\end{align}

\begin{align}\label{unit_formula3}
	|\rm Re(a_0)| < \frac{1}{2}.
\end{align}

\begin{align}\label{unit_formula4}
	a_0 = \frac{1}{16\pi\lambda}\left| \int\limits^{t_{+}}_{t_{-}}dt\cdot A \right |
\end{align}
where $\lambda$ is the kinematic function of the triangle and $A$ is the amplitude of the process.

Effective four-fermion operators are obtained by functional integration over massive modes of intermediate vector fields. A four-particle vertex can be formed from three-particle vertices of the interaction of two fermions with an auxiliary massive vector field and propagators of this massive field. In this case, the denominator of the auxiliary field propagator is replaced by a constant equal to the scale of the new physics.
The effective vertex of the interaction of right-handed polarized fermions with the auxiliary vector field has the form $\gamma(1+\gamma_5)/\sqrt{2}$, and that of left-handed polarized fermions - $\gamma(1-\gamma_5)/\sqrt{2 }$.

Using the Weyl representation for spinors and the method of helicity amplitudes, for each case of an anomalous operator, the amplitudes $2\to 2$ of $tt\to tt$ and $t\bar{t}\to t\bar{t}$ processes were calculated. It should be noted that the $tt\to tt$ process includes t and u channel components, and the $t\bar{t}\to t\bar{t}$ process includes s and t channel components, which were taken into account in the calculation. For each process, out of 16 possible helicity amplitudes, the one that gives the greatest contribution was chosen. The resulting helicity amplitudes were expressed in terms of the Mandelshtam invariant variables and integrated over the two-particle phase volume to calculate the corresponding partial amplitudes. 

Now let's take a closer look at each of the listed processes. Thus, the amplitude of the $tt\to tt$ process caused by the operator $O_{tt}^{1}$ is formed by two interacting right-handed fermionic currents and consists of one t-channel and one u-channel component. For simplicity, we fixed the angle $\phi=0$. The helicity amplitudes of this process do not depend on the angle $\theta$. To evaluate the applicability of the $O_{tt}^{1}$ operator, we chose one of the dominant helicity amplitudes, which is proportional to the square of the invariant mass s of the two top quarks. The remaining helicity amplitudes are proportional to the t-quark mass, and their relative contribution decreases with increasing s. Analytical expressions for the dominant helicity amplitude and the partial amplitude calculated on its basis are given in the Appendix in formulas (\ref{unit_formula5}) and (\ref{unit_formula6}), respectively. It can be seen that the partial amplitude of the process is proportional to the factor $(1+\beta)^2$, which tends to 4 for large s.

The process $tt\to tt$ with the operator $O_{QQ}^{1}$ is caused by the interaction of two left-handed currents and its dominant helicity amplitude looks similar to the case of the operator $O_{tt}^{1}$. The corresponding expressions for the amplitudes are given in formulas (\ref{unit_formula8}) and (\ref{unit_formula9}).

For a process with the operator $O_{Qt}^{1}$ the picture becomes more complicated. This process is caused by the interaction of the left and right currents of massive fermions. To the t-channel and u-channel amplitude components, two more are added, in which the right-handed and left-handed three-particle vertices are swapped. The dominant helicity amplitude of the process still does not depend on the angle $\theta$ and is directly proportional to s (formula \ref{unit_formula10}), however, due to a different combination of initial and final polarizations of fermions, as well as a different composition of the diagrams, the common factor of the dominant helicity amplitude is now proportional to $(1+\beta^2)$. This factor tends to 2 as s increases, therefore the partial amplitude of the process $tt\to tt$ with the operator $O_{Qt}^{1}$ (formula \ref{unit_formula11}) is half the corresponding amplitude with the operators $O_{ tt}^{1}$ or $O_{QQ}^{1}$.

Now consider the process $tt\to tt$ with the operators $O_{QQ}^{8}$ and $O_{Qt}^{8}$. The Lorentz structure of the corresponding amplitudes (formulas \ref{unit_formula11} and \ref{unit_formula14}) is similar to the cases of the operators $O_{QQ}^{1}$ and $O_{Qt}^{1}$, however, color factors $\lambda^a_{j,i}\lambda^a_{k,l}/4$ additionally appear in the diagrams.  In order to estimate the level of maximum influence of the operators $O_{QQ}^{8}$ and $O_{Qt}^{8}$, we took the largest possible value of the color factor $\lambda^a_{j,i}\lambda^a_{k, l}/4 = 1/2$. Thus, the obtained partial amplitudes for the process $tt\to tt$ with the operators $O_{QQ}^{8}$ and $O_{Qt}^{8}$ are two times smaller than the corresponding amplitudes with the operators $O_{QQ} ^{1}$ and $O_{Qt}^{1}$.

The listed operators also contribute to the processes $t\bar{t}\to t\bar{t}$. The amplitudes of this process include s-channel and t-channel components. The dominant helicity amplitude of the process $t\bar{t}\to t\bar{t}$ with the participation of the operator $O_{tt}^{1}$ (formula \ref{unit_formula16}) has an angular dependence and is proportional to the factor $-( 1+\cos{\theta})$, which is proportional to the Mandelstam variable u. Also, this helicity amplitude is proportional to the factor $(1+\beta)$, which tends to 2 as the value of s increases. Technically, the difference in the dominant helicity amplitudes for the processes $tt\to tt$ and $t\bar{t}\to t\bar{t}$ is due to different redistribution of momenta between the initial and final states, as well as different combinations of initial and final polarizations. These differences lead to the fact that at large s the partial amplitudes for the processes $tt\to tt$ with the operators $O_{tt}^{1}$ and $O_{QQ}^{1}$ are 4 times larger than the corresponding partial amplitudes for processes $t\bar{t}\to t\bar{t}$. At the same time, for large values of s, partial amplitudes with the operator $O_{Qt}^{1}$ for the process $t\bar{t}\to t\bar{t}$ coincide with the corresponding partial amplitude of the process $t\bar{t}\to t\bar{t}$.
	
From the calculations performed, it is clear that the partial amplitudes for the processes $t\bar{t}\to t\bar{t}$ do not exceed the partial amplitudes for the processes $tt\to tt$, therefore, further, to estimate the limits of applicability of the operators, we will use only partial amplitudes for $tt\to tt$ processes.

It should be noted that the considered helicity dominant amplitudes do not allow us to correctly estimate the contribution of processes with different operators. For such a comparison, it is necessary to calculate the full square of the amplitude modulus based on all the helicity amplitudes of the process. We used the dominant helicity amplitudes only to estimate the limits of applicability of the corresponding anomalous operators. We also note that the use of pair production processes of top quarks to determine the unitary limit of applicability of four-fermion operators provides a more conservative estimate than directly using the processes of production of three or four top quarks. We extend the values of the unitary limits obtained in the processes $tt\to tt$ to all processes involving the four-fermion operators under study. We used the obtained restrictions on the invariant mass of a pair of top quarks to set cutoffs on the invariant mass of three and four top quarks in the corresponding processes.

Based on the calculated partial amplitudes for the operators under study, we plotted graphs of the boundary of perturbative unitarity, where the invariant mass of a pair of top quarks is plotted along the x-axis, and the current upper experimental limits for the Wilson coefficients of anomalous operators are plotted along the y-axis. The intersection of the dotted lines corresponding to the accuracy of the Wilson coefficient measurements with the line of the unitarity boundary shows the value of the invariant mass of a pair of top quarks above which the effective field theory approach does not work. As experimental limits for the Wilson coefficients, their current values of measurement accuracy in the processes of production of three top quarks are taken.

\begin{figure}[!h!]
	\centering
	\includegraphics[width=0.4\textwidth,clip]{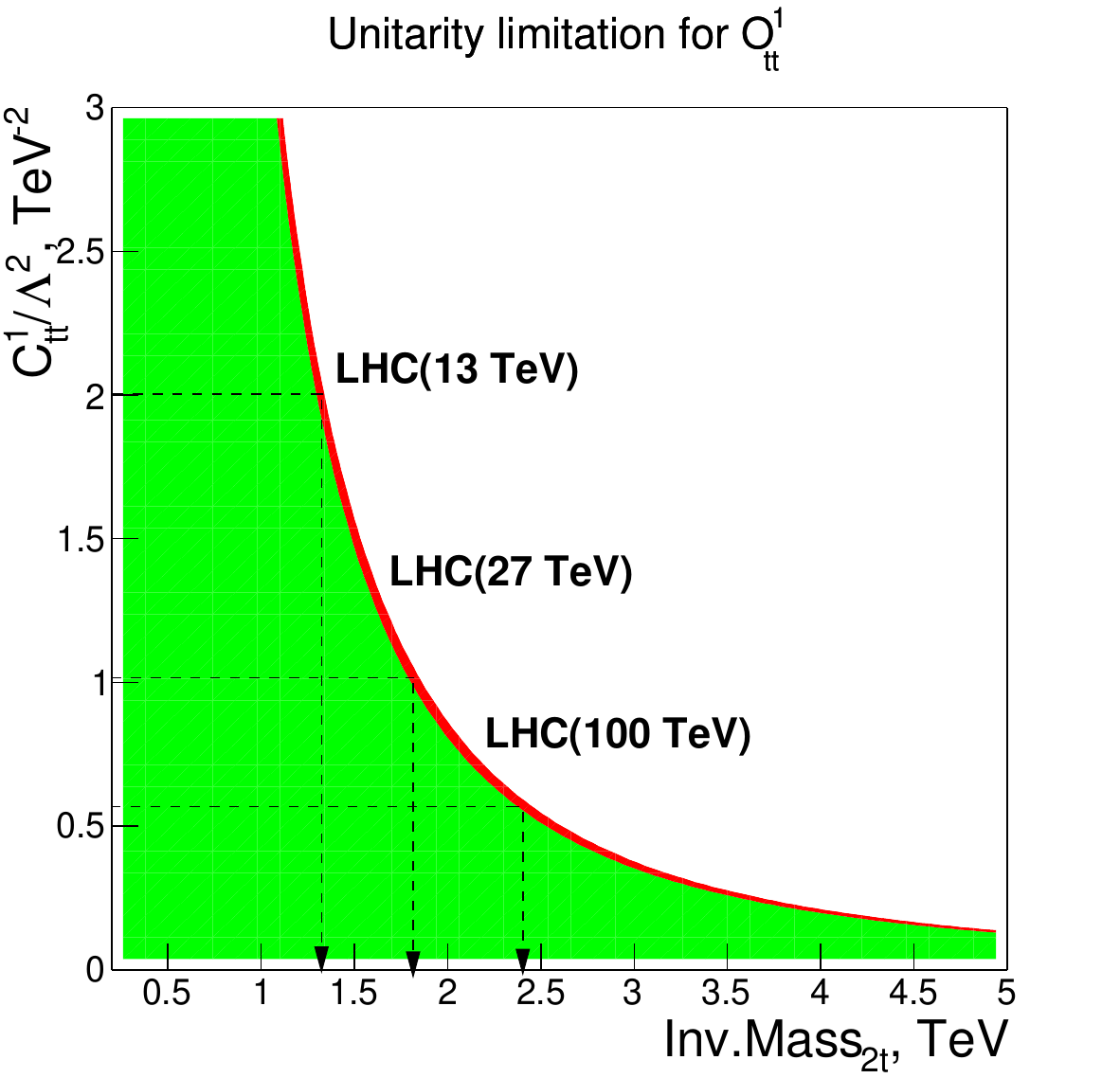}
	\includegraphics[width=0.4\textwidth,clip]{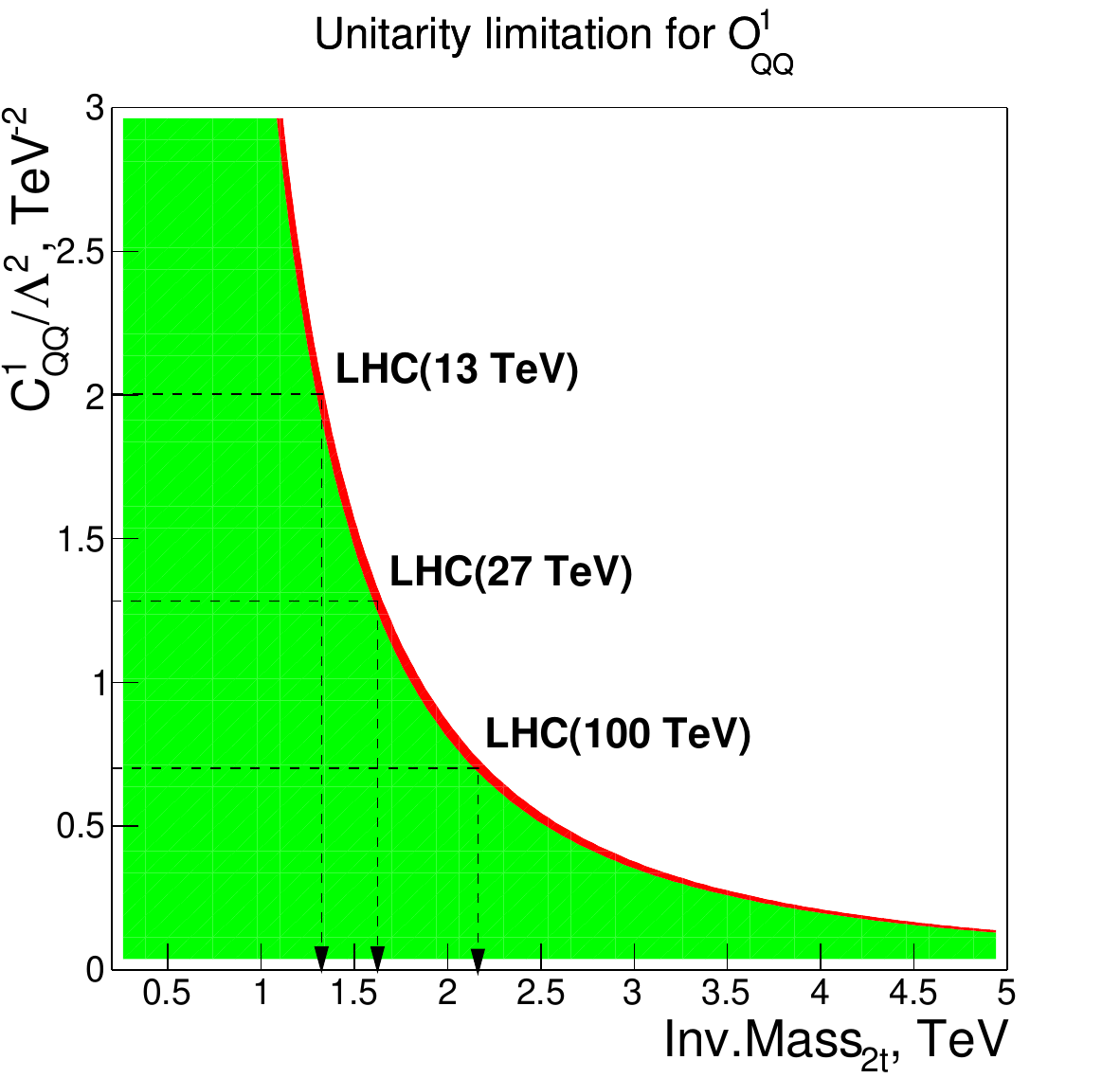}
	\includegraphics[width=0.4\textwidth,clip]{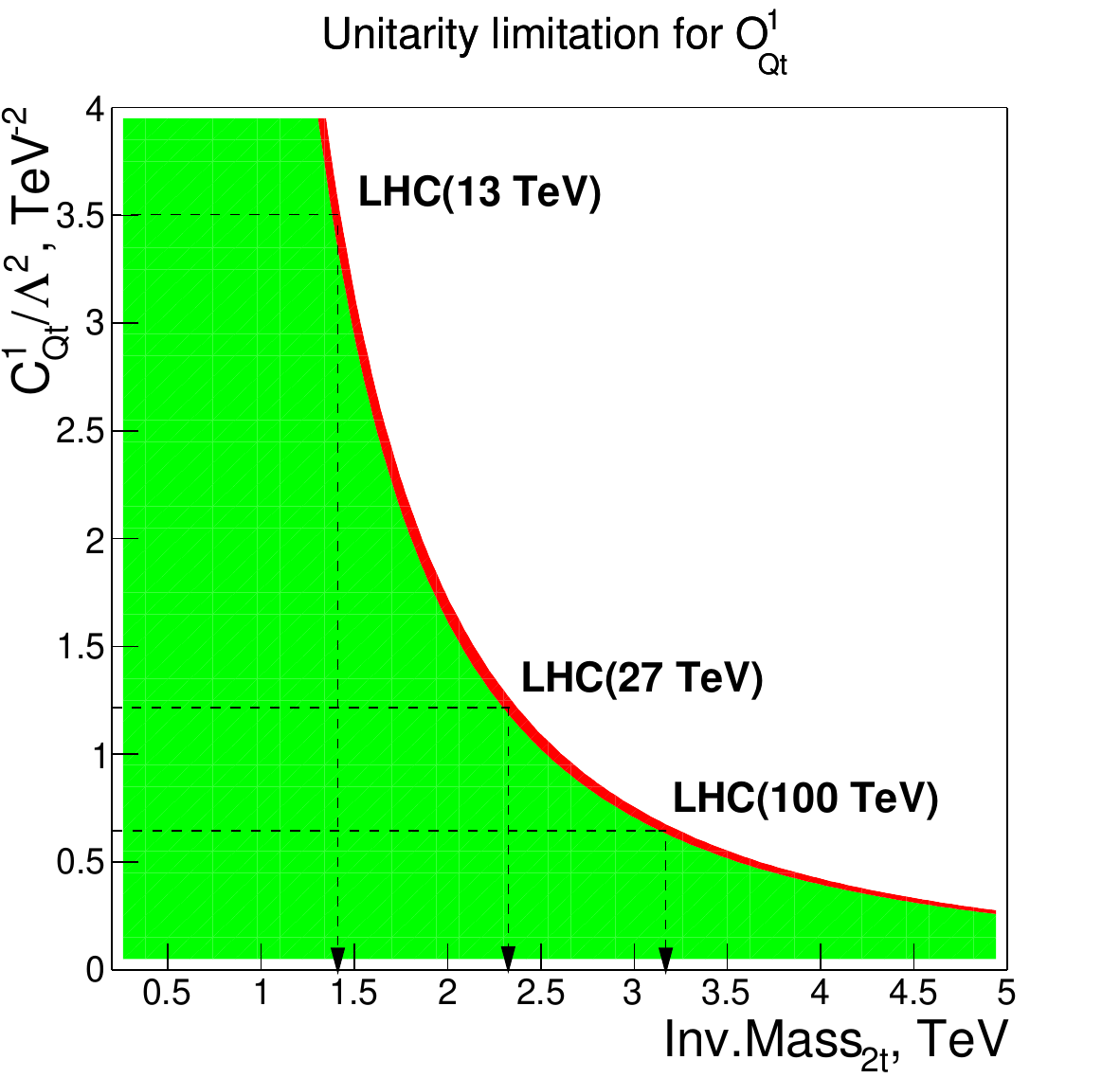}
	\includegraphics[width=0.4\textwidth,clip]{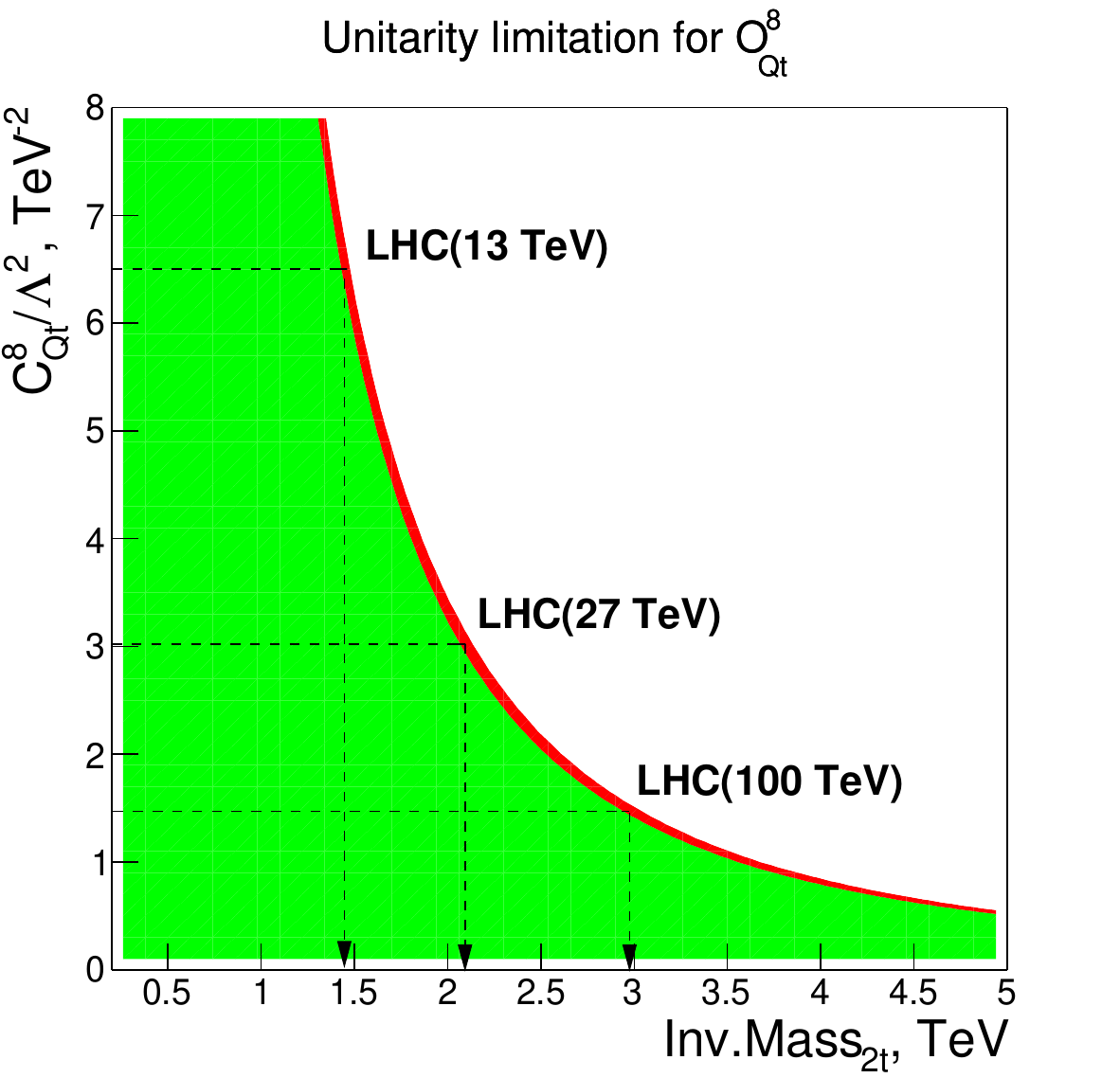}
	\includegraphics[width=0.4\textwidth,clip]{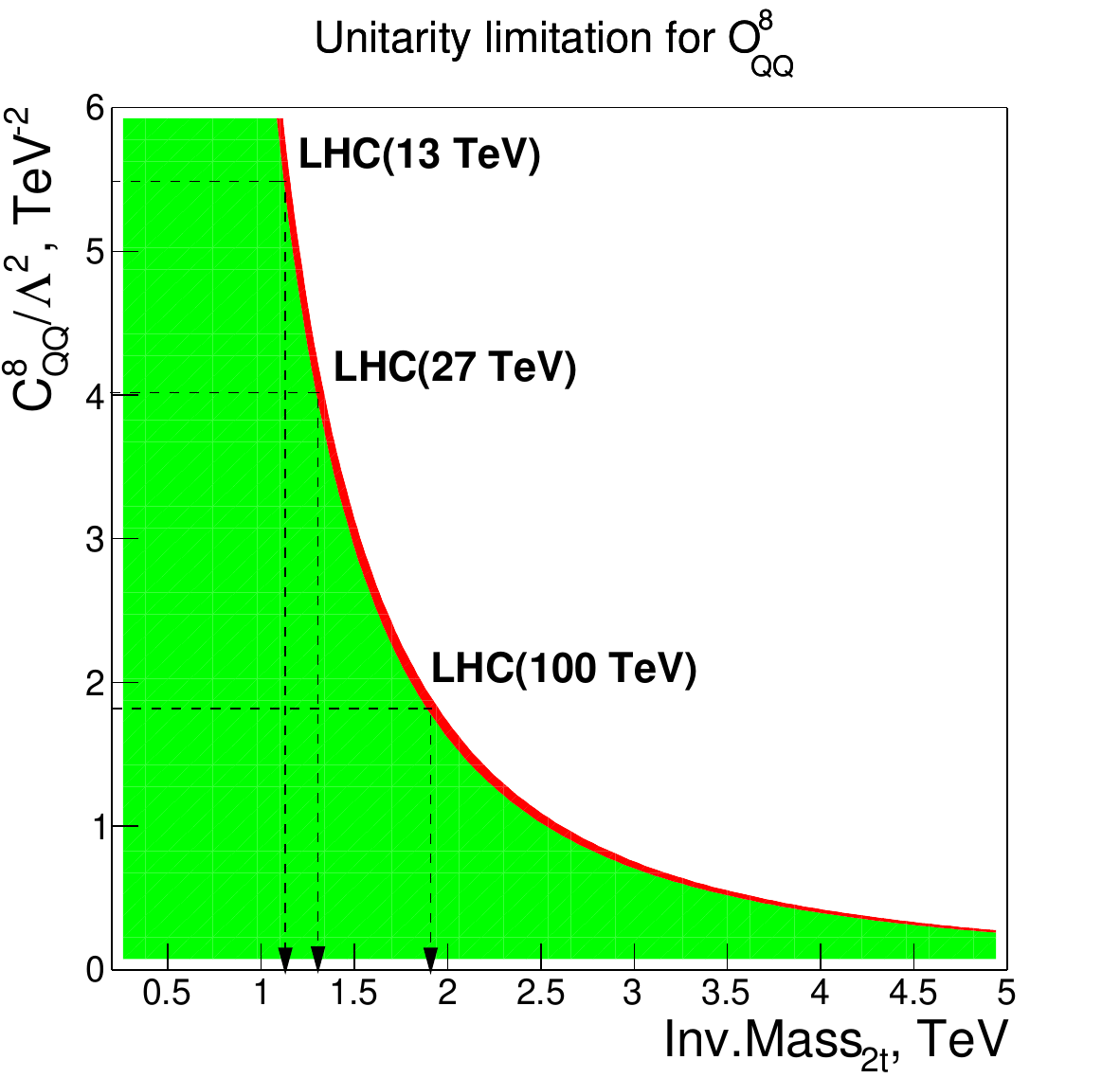}
	\caption{Perturbative unitarity limit $a_0=\frac{1}{2}$(red line) for various anomalous operators. Upper left: $O^{1}_{tt}$, upper right: $O^{1}_{QQ}$, middle left: $O^{1}_{Qt}$, middle right: $O^{8}_{Qt}$, bottom: $O^{8}_{QQ}$. The green zone corresponds to the allowed area. The dashed horizontal line indicates the experimental limits on the corresponding Wilson coefficients for different LHC energy values. Dashed vertical lines with an arrow indicate the limits of the invariant mass of two t-quarks for corresponding LHC energy values.}
	\label{fig_limit}      
\end{figure}

In Fig.~\ref{fig_limit}, the red color marks the values at which the partial amplitude $a_0 = \frac{1}{2}$. The green color marks the zone of parameters allowed from the point of view of perturbative unitarity. The dashed horizontal line indicates the experimental limits on the corresponding Wilson coefficients for different LHC operating modes. Dashed vertical lines with an arrow indicate the limits of the invariant mass of two t-quarks for corresponding LHC modes. 
When evaluating the unitarity limitation on the operators $O^{1}_{tt}$, $O^{1}_{QQ}$, $O^{1}_{Qt}$, $O^{8}_{Qt}$ at the LHC energy equal 13 TeV, the current values of the measurement accuracy of the Wilson coefficients obtained in the CMS experiment \cite{CMS:2019jsc} were used. At the same time, when estimating the corresponding limitations for the LHC energy equal to 27 and 100 TeV, the expected values of the measurement accuracy, obtained theoretically, were used. For operator $O^{8}_{QQ}$ only theoretical values of measurement accuracy were used.

From Fig.~\ref{fig_limit} it is clear that the lower the value of the upper experimental limit on the Wilson coefficient of the operator, the higher the value of the invariant mass of the process is allowed by the condition of perturbative unitarity. At the same time, the upper limit of the Wilson coefficient of the operator depends on the cross section of the process under study. The larger the cross section of a hypothetical process involving an anomalous operator, the greater the accuracy of its measurement is allowed. In the absence of manifestations of new physics, this leads to lower values of the experimental limits on the value of the Wilson coefficient.

For example, consider the behavior of unitary limits for 4t operators at the LHC 13 TeV mode.
(Fig.~\ref{fig_limit} top left) shows the unitary bound for the operator $O^{1}_{tt}$. With an experimental limit on the interaction parameter of this operator equal to 2 $TeV^{-2}$, we obtain an upper unitary limit on the invariant mass of the process equal to 1.5 TeV. The same values are obtained for the operator $O^{1}_{QQ}$ on (Fig.~\ref{fig_limit} upper right). For the operator $O^{1}_{Qt}$ (Fig.~\ref{fig_limit} middle left) the picture changes. Since the partial amplitude for this operator is two times less than for the operators $O^{1}_{tt}$ and $O^{1}_{QQ}$, the line of the unitary boundary moves further from the coordinate axes. At the same time, the cross section for processes with four top quarks involving this operator is smaller than the corresponding cross sections for the operators $O^{1}_{tt}$ and $O^{1}_{QQ}$, therefore the accuracy of the experimental measurement of $C^{1}_{Qt}/\Lambda^2$ is more rough and equal to 3.5 TeV$^{-2}$. It turns out that the intersection of the rough value of the experimental limit with the moved unitary boundary again gives a unitary limit on the invariant mass equal to 1.5 TeV. This trend continues for the $O^{8}_{Qt}$ operator (Fig.~\ref{fig_limit} middle right). The partial amplitude corresponding to this operator is additionally two times smaller, and the measurement accuracy $C^{8}_{Qt}/\Lambda^2$ is even rougher and is equal to 6.5 TeV$^{-2}$. This leads to the fact that the unitary constraint on the invariant mass of a process with this operator is again around the value of 1.5 TeV. Thus, the total unitary limit for all considered 4t operators in the LHC 13 TeV mode is close to the value of 1.5 TeV.

It is also clear from (Fig.~\ref{fig_limit}) that although changing the LHC operating mode from 13 TeV to 27 TeV and 100 TeV leads to an expansion of the unitary limits of operators, this expansion is not very significant and is equal to 2 TeV and 3 TeV, respectively. This situation is due to the fact that the cross sections for processes with four top quarks are accumulated at invariant masses not exceeding 10 TeV.
 
It should be noted that in the case of the considered 4-fermion operators, the partial amplitude grows proportionally to s, which limits the use of these operators already for the invariant mass of two top quarks equal to 1–3 TeV.
For comparison, the unitary limitation obtained for single top production with anomalous operators contributing to the Wb \cite{Boos:2023slg} vertex, as well as with FCNC operators \cite{Boos:2020kqq}, is at the level of 10 TeV and higher.	

\section{Setting limits on Wilson coefficients}
\label{sec:limits}

\begin{table}[b]
    \centering
    \begin{tabular}{l l|c c| c c}
        ~ & ~ & \multicolumn{2}{c}{$pp \rightarrow 4 top$} & \multicolumn{2}{c}{$pp \rightarrow 3 top + X$} \\
        Energy &
        int. Luminosity &
         $\delta sys$, \%  &  
         $\delta stat$, \%  &
         $\delta sys$, \%  &  
         $\delta stat$, \%  \\ \hline
        13 TeV & 138 fb$^{-1}$ & 13 &  24 & 13 & 76 \\
        14 TeV & 3 ab$^{-1}$ &  13 &  5 &  13 & 14\\
        27 TeV & 15 ab$^{-1}$ &  13 &  0.6 &  13 &  2 \\
        100 TeV & 25 ab$^{-1}$ &  13 &  0.1 &  13 &  0.3 \\
    \end{tabular}
    \caption{Estimated experimental systematic and statistical uncertainties for processes of three and four top quark production at energies of 13, 14, 27 and 100 TeV}
    \label{Tab:exp_uncert}
\end{table}

Having the values of coefficients $\sigma^{(1)}$ and $\sigma^{(2)}$ and their uncertainties, calculating constraints on $c_i$ is a matter of choice of a suitable statistical model. In the statistical model, we compare the calculated with Eq.~\ref{sigma_eq} total cross section with the measured or SM expected cross section, with corresponding uncertainties. We consider three types of uncertainties. The first type is theory uncertainty which has been discussed in Sec.~\ref{Sec:cs}, the numbers are provided in Tables~\ref{t1},~\ref{t2}. The other two uncertainties are experimental systematic and statistical uncertainties. Based on the recent measurements of four top-quark production in CMS~\cite{CMS:2023ftu} and ATLAS~\cite{ATLAS:2023ajo} experiments one can extrapolate the statistical uncertainty and estimate experimental systematic relative uncertainty to be the same for all calculations. This approach is more or less conservative since the analysis methodology usually improves and one can expect smaller systematic uncertainties in the future. Since the number of expected events $n\sim \sigma L$ the extrapolation of statistical relative uncertainty can be taken as $\delta(n1)/\delta(n2)=\sqrt{(\sigma_2 L_2)/(\sigma_1 L_1)}$. Based on the CMS results~\cite{CMS:2023ftu} $\sigma_{\rm 4top}=17.7\rm ^{+3.7}_{-3.5} (stat) ^{+2.3}_{-1.9} (sys)$ fb we estimate experimental uncertainties as listed in Table~\ref{Tab:exp_uncert}. 
The integrated luminosity (L) for the calculations has been taken as 138 fb$^{-1}$ for 13 TeV (available experimental results), 3 ab$^{-1}$ for 14 TeV (HL-LHC), 15 ab$^{-1}$ for 27 TeV (HE-LHC) and 25 ab$^{-1}$ for 100 TeV (FCC).

\subsection{Cross checks of the methodology}

\begin{table}[t]
    \centering
    \begin{tabular}{l | c c c c c}
Stat. model & $C_{tt}^1$ & $C_{QQ}^1$ & $C_{Qt}^1$ & $C_{Qt}^8$ & $C_{QQ}^8$ \\ \hline\hline
$\chi^2$,1D,4t & [-27,27] & [-39,39] & - & [-28,28] & - \\
EFTfitter,1D,4t & [-27,27] & [-39,39] & - & [-28,28] & [-130,130] \\
SMEFiT,1D,4t & [-27,27] & [-39,39] & - & [-28,28] & [-130,130] \\ \hline
$\chi^2$,1D,3t & -  & [-27,27] & [-64,64] & [-123,121] & [-43,43] \\
EFTfitter,1D,3t & - & [-27,27] & [-64,64] & [-123,121] & [-43,43] \\
SMEFiT,1D,3t & -    & [-27,27] & [-64,64] & [-123,121] & [-43,43] \\ \hline
$\chi^2$,1D,3+4t  & [-27,27] & [-22,22] & [-64,64] & [-27,27] & [-42,42] \\
EFTfitter,1D,3+4t & [-27,27] & [-22,22] & [-64,64] & [-27,27] & [-42,42] \\
SMEFiT,1D,3+4t    & [-27,27] & [-22,22] & [-64,64] & [-27,27] & [-42,42] \\ \hline
EFTfitter,5D,4t & [-138,138] & [-141,141] & - & [-138,138] & - \\
SMEFiT,5D,4t    & [-138,138] & [-141,141] & - & [-138,138] & - \\ \hline
EFTfitter,5D,3t & - & [-125,125] & - & - & - \\
SMEFiT,5D,3t    & - & [-125,125] & - & - & - \\ \hline
EFTfitter,5D,3+4t & [-132,132] & [-123,123] & - & [-140,140] & - \\
SMEFiT,5D,3+4t    & [-132,132] & [-123,123] & - & [-140,140] & - \\ \hline    \end{tabular}
    \caption{Comparison of the expected limits on  $C_k/\Lambda^2 \rm [TeV^{-2}]$ estimated for $pp \rightarrow 4 top$ (4t) and $pp \rightarrow 3 top + X$ (3t) cross sections with only linear $\sigma^{(1)}$ EFT terms. The statistical models with only one Wilson coefficient have marked 1D and the models with all five coefficients considered simultaneously have marked 5D. The models with the combination of 4t and 3t processes are marked 3+4t. The results were obtained under LHC conditions (13 TeV, 138 fb$^{-1}$).}
    \label{Tab:lin_limits}
\end{table}

For the first fit, a statistical model based on the chi-square distribution was used. Later, the results were validated using acknowledged EFTfitter~\cite{eft_fitter} and SMEFiT~\cite{Giani:2023gfq} packages. The first comparison between different statistical models was done for linear terms in Eq.~\ref{sigma_eq} where only $\sigma^{(1)}$ terms are taken into account. Based on the calculated cross sections of the four and three top-quarks production processes in SM and with EFT contribution one can estimate the expected results with different statistical approaches. The achieved with the fit 95\% CL exclusion limits on the Wilson coefficients $C_k/\Lambda^2 (TeV^{-2})$ are shown in Table~\ref{Tab:lin_limits}. 
Two processes $pp \rightarrow 4 top$ (4t) and $pp \rightarrow 3 top + X$ (3t) at $\sqrt{s}=13$ TeV are considered separately and as a combination in one statistical model. Two types of statistical models are realized. The 1D approach with a variation of only one Wilson coefficient in the statistical model, and the 5D approach with a simultaneous variation of all five coefficients in the same model. The combination of 4t and 3t processes has been considered in a dedicated statistical model marked as (3+4t).
For some of the couplings, there is no sensitivity (flat posterior distribution) with linear terms in the range from -150 to 150 for the value of the coefficient. In this case, the sign ``-'' is shown in the table. 

 The comparison in Table~\ref{Tab:lin_limits} for the linear EFT terms demonstrates significant improvement of the $C_{QQ}^1$, $C_{Qt}^1$ and  $C_{QQ}^8$ limits for the scenarios where triple top-quark production has taken into account. The limits in Table~\ref{Tab:lin_limits} demonstrate very weak sensitivity of the linear EFT terms to considered operators, the limits are far beyond the perturbative unitarity limits of $4\pi$.
 
\begin{figure}[!h!]
\centering
\includegraphics[width=0.47\textwidth,clip]{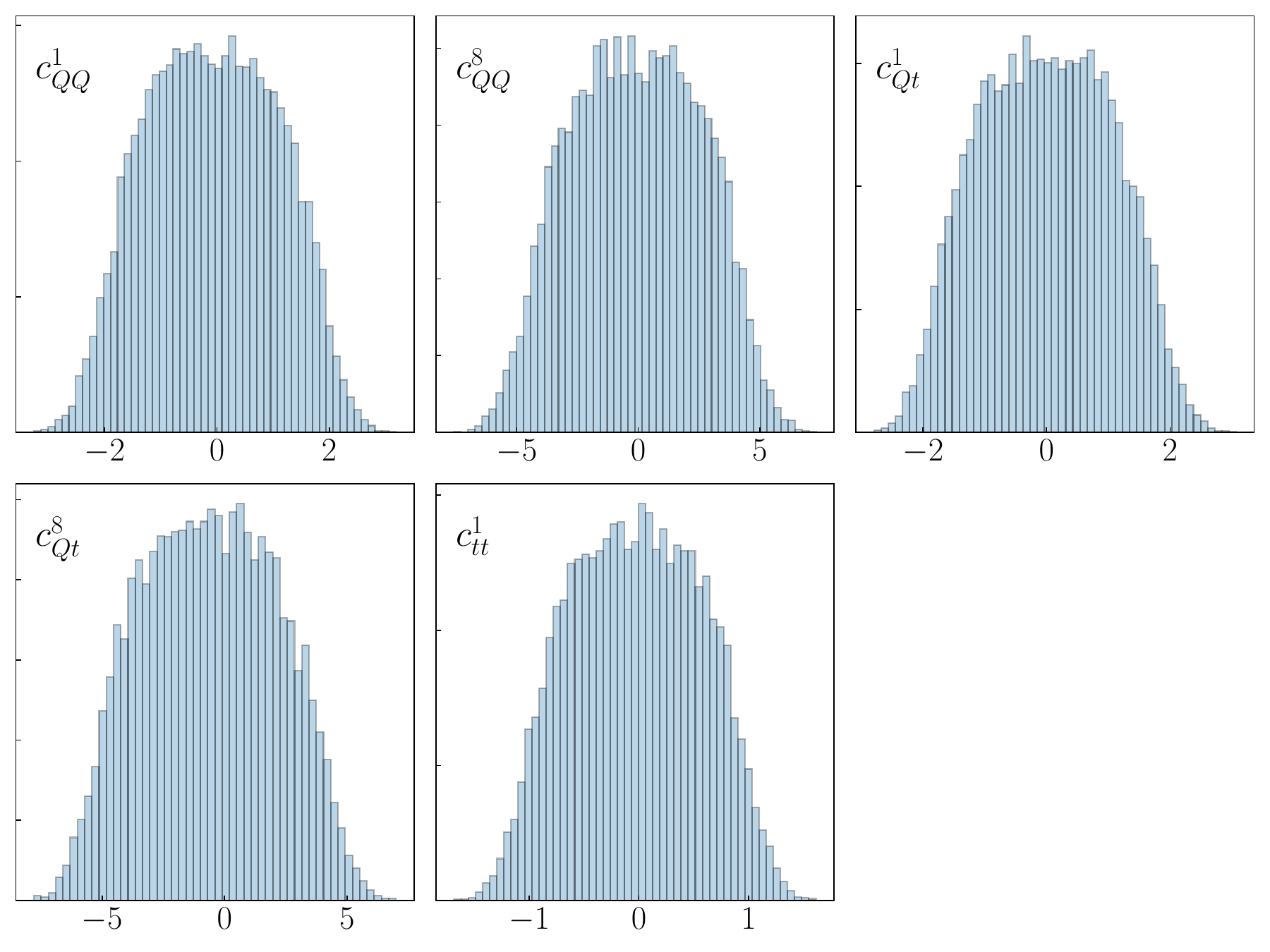}
\includegraphics[width=0.47\textwidth,clip]{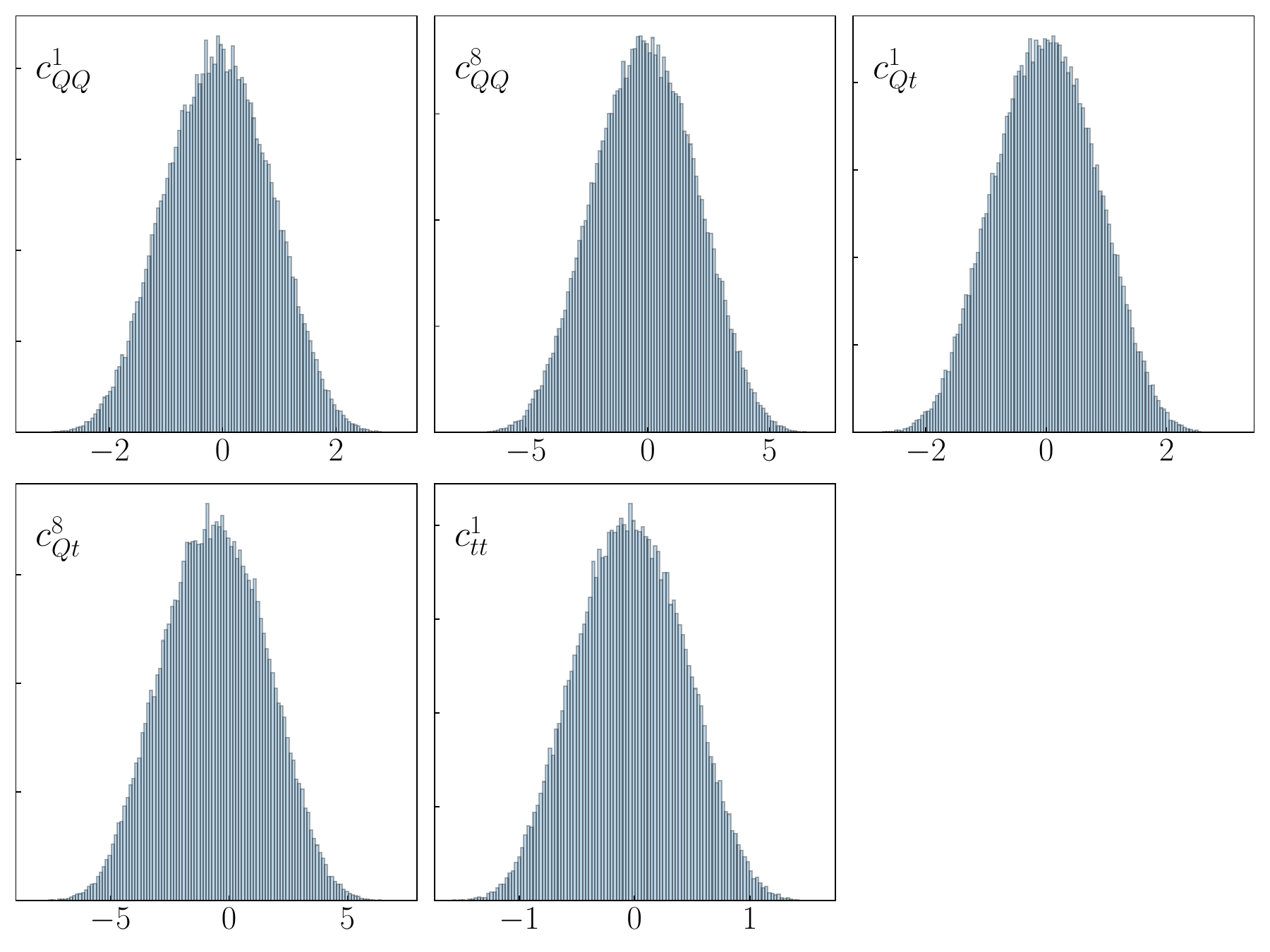} 
\caption{Posterior distributions of probability density function for the Wilson coefficients in 1D (left plots) and 5D (right plots) statistical models, in the scenario with quadratic terms are taken into account.}
\label{coeff_posterior}
\end{figure}

In the scenarios with quadratic EFT terms, where $\sigma^{(2)}$ in Eq.~\ref{sigma_eq} are taken into account, the limits are significantly tighter than with only linear terms scenarios. The corresponding limits with quadratic terms are shown in Table~\ref{Tab:quad_limits} for different statistical models at $\sqrt{s}=13$ TeV.
\begin{table}[t]
    \centering
    \begin{tabular}{l | c c c c c}
Stat. model & $C_{tt}^1$ & $C_{QQ}^1$ & $C_{Qt}^1$ & $C_{Qt}^8$ & $C_{QQ}^8$ \\ \hline\hline
$\chi^2$,1D,4t & [-1.3,1.2] & [-2.5,2.3] & [-2.2,2.2] & [-6.3,5.2] & [-5.6,5.5] \\
EFTfitter,1D,4t & [-1.1,1.1] & [-2.2,2.1] & [-2.0,2.0] & [-5.7,4.6] & [-5.1,4.9] \\
SMEFiT,1D,4t & [-1.1,1.1] & [-2.2,2.1] & [-2.0,2.0] & [-5.7,4.6] & [-5.0,4.8] \\ \hline
$\chi^2$,1D,3t & [-4.2,4.1]  & [-2.9,3.2] & [-2.9,3.0] & [-6.0,6.3] & [-5.8,6.8] \\
EFTfitter,1D,3t & [-3.8,3.7] & [-2.5,2.9] & [-2.6,2.7] & [-5.4,5.6] & [-5.2,6.1] \\
SMEFiT,1D,3t & [-3.7,3.7]  & [-2.5,2.9] & [-2.6,2.7] & [-5.3,5.6] & [-5.1,6.1] \\ \hline
$\chi^2$,1D,3+4t  & [-1.3,1.2] & [-2.2,2.2] & [-2.1,2.1] & [-5.2,4.7] & [-4.8,5.0] \\
EFTfitter,1D,3+4t & [-1.1,1.1] & [-2.0,2.0] & [-1.8,1.9] & [-4.7,4.2] & [-4.3,4.5] \\
SMEFiT,1D,3+4t    & [-1.1,1.0] & [-2.0,2.0] & [-1.8,1.8] & [-4.7,4.2] & [-4.2,4.5] \\ \hline
EFTfitter,5D,4t & [-0.94,0.92] & [-2.3,2.2] & [-1.7,1.6] & [-4.9,3.7] & [-5.2,5.1] \\
SMEFiT,5D,4t    & [-0.95,0.90] & [-1.8,1.7] & [-1.6,1.6] & [-4.8,3.6] & [-4.2,4.0] \\ \hline
EFTfitter,5D,3t & [-3.3,3.7] & [-2.1,2.4] & [-2.2,2.5] & [-4.6,5.4] & [-4.3,5.5] \\
SMEFiT,5D,3t    & [-3.1,3.0] & [-2.0,2.4] & [-2.1,2.2] & [-4.3,4.6] & [-4.2,5.1] \\ \hline
EFTfitter,5D,3+4t & [-0.98,0.92] & [-1.8,1.8] & [-1.5,1.6] & [-4.0,3.4] & [-3.9,4.0] \\
SMEFiT,5D,3+4t    & [-0.95,0.90] & [-1.6,1.6] & [-1.5,1.5] & [-4.0,3.3] & [-3.5,3.7] \\ \hline    \end{tabular}
    \caption{Comparison of the expected limits on  $C_k/\Lambda^2 \rm [TeV^{-2}]$ estimated for $pp \rightarrow 4 top$ (4t) and $pp \rightarrow 3 top + X$ (3t) cross sections with quadratic $\sigma^{(2)}$ EFT terms are taken into account. The statistical models with only one Wilson coefficient have marked 1D and the models with all five coefficients considered simultaneously have marked 5D. The models with the combination of 4t and 3t processes are marked 3+4t. The results were obtained under LHC conditions (13 TeV, 138 fb$^{-1}$).}
    \label{Tab:quad_limits}
\end{table}
The Tables~\ref{Tab:lin_limits},~\ref{Tab:quad_limits} demonstrate good agreement between SMEFiT, EFTfitter and $\chi^2$ results. For the quadratic fit, the $\chi^2$ test provides a bit wider limits. For the linear fit multi-dimensional variation leads to much weaker limits than the one-dimensional statistical model which is comparable with previous results with linear fit~\cite{Ellis:2020unq}. In the scenario with quadratic EFT terms, the multi-dimensional statistical model leads to tighter limits than the one-dimensional statistical model. Such behavior is comparable with previous results for the quadratic fits~\cite{Ethier:2021bye}. In Fig.~\ref{coeff_posterior} the posterior distributions of the probability density function for the Wilson coefficients are provided for the one-dimensional (left plots) and multi-dimensional (right plots) statistical models, such example has taken for four-top-quark production at $\sqrt{s}=13$ TeV with quadratic EFT terms. The distributions are calculated with the SMEFiT package. Due to quadratic terms, the distributions in multi-dimensional models are sharper and the integral for 0.95 quantile is gaining faster. Depending on the uncertainties such behavior can give tighter limits in multi-dimensional statistical model than in one-dimensional. 

The achieved upper limits in Table~\ref{Tab:quad_limits} for the scenarios with quadratic EFT terms demonstrate rather similar sensitivity of four- and triple-top quark production processes to the considered EFT operators. The combination of four- and triple-top quark production in one statistical model leads to better sensitivity, in general.
    
The theoretical constraints seem to coincide pretty well with experimental limits obtained from four top-quark production in CMS~\cite{CMS:2019jsc} and ATLAS~\cite{ATLAS:2023ajo} experiments. Since there is a good agreement in the results from EFTfitter and SMEFiT packages, EFT limits in the next sections are shown only from the SMEFiT package.

\subsection{Electroweak and QCD contributions in triple top-quark production}

\begin{table}[b]
    \centering
    \begin{tabular}{l | c c c c c}
model & $C_{tt}^1$ & $C_{QQ}^1$ & $C_{Qt}^1$ & $C_{Qt}^8$ & $C_{QQ}^8$ \\ \hline\hline
3t,QCD,1D & [-3.6,3.7] & [-2.6,2.5] & [-2.5,2.5] & [-6.3,4.8] & [-6.5,5.3] \\
3t,QCD+EW,1D & [-3.4,3.4] & [-2.3,2.7] & [-2.4,2.5] & [-5.0,5.2] & [-4.8,5.6] \\ \hline
3t,QCD,5D & [-3.0,3.0] & [-2.2,2.1] & [-2.0,2.1] & [-5.3,3.9] & [-5.5,4.2] \\
3t,QCD+EW,5D & [-2.8,2.8] & [-1.9,2.2] & [-1.9,2.1] & [-4.1,4.3] & [-3.8,4.7] \\
 \end{tabular}
    \caption{Comparison of the expected limits on  $C_k/\Lambda^2 \rm [TeV^{-2}]$ estimated for $pp \rightarrow 3 top + X$ (3t) cross sections with quadratic $\sigma^{(2)}$ EFT terms. Two simulation models are considered, the only QCD diagrams and QCD plus EW diagrams and their interference. The statistical models with only one Wilson coefficient have marked 1D and the models with all five coefficients considered simultaneously have marked 5D. The results were obtained under LHC conditions (13 TeV, 138 fb$^{-1}$).}
    \label{Tab:limits_ew-qcd-3t}
\end{table}

The importance of electroweak contribution (EW) to the triple top quark production processes has been shown in the article~\cite{Boos:2021yat}. The EW diagrams have the same rate as diagrams with a gluon (QCD), also the interference between EW and QCD diagrams is of almost the same rate as QCD or EW contribution and is negative. If one takes into account only QCD diagrams (e.g. it is the default setting in MadGraph) the total cross section will be almost correct due to the cancellation of negative interference terms and EW contribution, but the kinematic properties can be significantly different. In this subsection, we compare how the inclusion of EW contribution changes the EFT limits in comparison with the limits where only QCD contribution has been taken into account. The EFT limits shown in Table~\ref{Tab:limits_ew-qcd-3t} are calculated with triple top quark production processes with only QCD diagrams (marked as QCD) and in the complete case with QCD, EW diagrams and their interference (marked as QCD+EW). Two statistical models are shown with 1D and 5D variations. The quadratic EFT terms with $\sigma^{(2)}$ terms are taken into account. 

Direct comparison of the expected limits in Table~\ref{Tab:limits_ew-qcd-3t} demonstrate notable improvement in the sensitivity for the calculations with the correct simulation of all QCD, EW and interference contributions.

\subsection{Limits with unitarity bound cuts}
The unitarity bounds considered in Sec.~\ref{unitarity} have to be taken into account with additional requirements for an invariant mass of all top-quark pairs. Such additional requirements have been applied for the calculation of cross sections and $\sigma^{(1)}$, $\sigma^{(2)}$ EFT terms. The achieved limits on  $C_k/\Lambda^2 (TeV^{-2})$ are provided in Table~\ref{Tab:limits_unitarity} and are marked as (cut) when the requirement has applied, for the comparison of the results without unitarity bound requirements (nocut) are also shown. 
\begin{table}[t]
    \centering
    \begin{tabular}{l | c c c c c}
model & $C_{tt}^1$ & $C_{QQ}^1$ & $C_{Qt}^1$ & $C_{Qt}^8$ & $C_{QQ}^8$ \\ \hline\hline
4t,nocut,1D & [-1.1,1.1] & [-2.2,2.1] & [-2.0,2.0] & [-5.7,4.6] & [-5.0,4.8] \\
4t,cut,1D & [-1.2,1.2] & [-2.4,2.3] & [-2.2,2.2] & [-6.8,5.0] & [-6.0,5.7] \\ \hline
3t,nocut,1D &  [-3.7,3.7]  & [-2.5,2.9] & [-2.6,2.7] & [-5.3,5.6] & [-5.1,6.1] \\
3t,cut,1D & [-4.3,4.2] & [-2.9,3.2] & [-3.1,3.2] & [-6.9,7.3] & [-6.4,7.7] \\\hline
3+4t,nocut,1D & [-1.1,1.0] & [-2.0,2.0] & [-1.8,1.8] & [-4.7,4.2] & [-4.2,4.5] \\
3+4t,cut,1D & [-1.2,1.2] & [-2.2,2.2] & [-2.1,2.1] & [-5.8,4.8] & [-5.2,5.4] \\ \hline
4t,nocut,5D & [-0.95,0.90] & [-1.8,1.7] & [-1.6,1.6] & [-4.8,3.6] & [-4.2,4.0] \\
4t,cut,5D & [-1.0,1.0] & [-2.0,1.9] & [-1.8,1.9] & [-5.7,4.1] & [-4.6,4.4] \\ \hline
3t,nocut,5D & [-3.1,3.0] & [-2.0,2.4] & [-2.1,2.2] & [-4.3,4.6] & [-4.2,5.1] \\
3t,cut,5D & [-3.5,3.4] & [-2.3,2.7] & [-2.5,2.7] & [-5.6,6.1] & [-5.1,6.5] \\\hline
3+4t,nocut,5D & [-0.95,0.90] & [-1.6,1.6] & [-1.5,1.5] & [-4.0,3.3] & [-3.5,3.7] \\
3+4t,cut,5D & [-1.0,1.0] & [-1.8,1.8] & [-1.7,1.7] & [-4.8,3.8] & [-4.1,4.3] \\ \end{tabular}
    \caption{Comparison of the expected limits on  $C_k/\Lambda^2 \rm [TeV^{-2}]$ estimated for  $pp \rightarrow 4 top$ (4t) and $pp \rightarrow 3 top + X$ (3t) cross sections with quadratic $\sigma^{(2)}$ EFT terms. The limits are shown for the case of unitarity bound cuts applied at the simulation level (cut) and without such cuts (nocut). The statistical models with only one Wilson coefficient have marked 1D and the models with all five coefficients considered simultaneously have marked 5D. The results were obtained under LHC conditions (13 TeV, 138 fb$^{-1}$).}
    \label{Tab:limits_unitarity}
\end{table}

Expected limits in Table~\ref{Tab:limits_unitarity} achieved with additional unitarity bound cuts obviously worse than without such cuts, but not significantly. Since the unitarity is necessary requirement the simulation for the final results in the next section obeys the calculated in Sec.~\ref{unitarity} unitarity bound cuts.

The general comparison of the separate and combined results for the four and three top quark production processes demonstrates significant improvement in sensitivity for the linear EFT terms, especially for $O_{QQ}^1$, $O_{Qt}^1$ and  $O_{QQ}^8$ operators.
In the scenarios with quadratic EFT terms the Wilson coefficient of the operator $O_{tt}^1$ is better constrained by the process of four top quark production. However, in the case of other operators, the three top quark production apparently provides similar limits. The combination of four and three top quark production processes gains in sensitivity for most of the considered operators. 

\section{Summary}

\begin{table}[t]
    \centering
    \begin{tabular}{l | c c c c c}
Energy, model & $C_{tt}^1$ & $C_{QQ}^1$ & $C_{Qt}^1$ & $C_{Qt}^8$ & $C_{QQ}^8$ \\ \hline\hline
13 TeV, 4t & [-1.2, 1.2] & [-2.4, 2.3] & [-2.2, 2.2] & [-6.8, 5.0] & [-6.0, 5.7] \\
13 TeV, 3t & [-4.3, 4.2] & [-2.9, 3.2] & [-3.1, 3.2] & [-6.9, 7.3] & [-6.4, 7.7] \\
13 TeV, 3+4t & [-1.2, 1.2] & [-2.2, 2.2] & [-2.1, 2.1] & [-5.8, 4.8] & [-5.2, 5.4] \\ \hline
14 TeV, 4t &[-1.1, 1.0] &[-2.1, 2.0] &[-1.9, 1.9] &[-5.8, 4.2] & [-5.2, 4.9] \\
14 TeV, 3t &[-2.5, 2.5] &[-1.6, 2.0] &[-1.8, 1.9] &[-3.9, 4.4] &[-3.7, 5.1] \\
14 TeV, 3+4t &[-1.1, 1.0] &[-1.5, 1.7] &[-1.5, 1.6] &[-3.8, 3.6] &[-3.5, 4.3] \\ \hline
27 TeV, 4t &[-0.90, 0.83] &[-1.7, 1.6] &[-1.6, 1.6] &[-4.9, 3.6] &[-4.4, 4.2] \\
27 TeV, 3t &[-2.0, 2.0] &[-1.3, 1.5] &[-1.4, 1.6] &[-3.3, 3.9] &[-2.7, 4.1] \\
27 TeV, 3+4t &[-0.88, 0.83] &[-1.2, 1.3] &[-1.3, 1.3] &[-3.2, 3.2] &[-2.6, 3.5] \\ \hline
100 TeV, 4t &[-0.68, 0.66] &[-1.3, 1.3] &[-1.2, 1.2] &[-3.8, 3.0] &[-3.7, 3.6] \\
100 TeV, 3t &[-1.3, 1.4] &[-0.89, 1.0] &[-1.0, 1.1] &[-2.1, 2.6] &[-1.8, 2.7] \\
100 TeV, 3+4t &[-0.67, 0.64] &[-0.85, 0.94] &[-0.93, 0.94] &[-2.1, 2.3] &[-1.8, 2.5] \\ \hline
\end{tabular}
    \caption{Comparison of the expected limits on  $C_k/\Lambda^2 \rm [TeV^{-2}]$ estimated for  $pp \rightarrow 4 top$ (4t) and $pp \rightarrow 3 top + X$ (3t) cross sections with quadratic $\sigma^{(2)}$ EFT terms. The limits are shown with applied unitarity bound requirements. The results have been achieved in a one-dimensional statistical model with a variation of each coefficient separately.}
    \label{Tab:summary}
\end{table}

In the present study, numerical simulations of processes of three and four top quark production were carried out. Simulation results are presented for both the Standard Model and dimension six SMEFT. For the latter case, the EFT validity issue was addressed by analyzing partial unitarity requirements for processes of interest, deriving corresponding kinematic cuts, and implementing them into the simulation. 

Theoretical constraints on the Wilson coefficients $C_k/\Lambda^2 (TeV^{-2})$ of respective SMEFT operators are obtained for both channels of three and four top quark production and are shown in Table~\ref{Tab:summary}. 

Analytical expressions for the partial amplitudes of the processes $tt\to tt$ and $t\bar{t}\to t\bar{t}$ caused by the operators $O^1_{tt}$, $O^1_{QQ}$, $O^1_{Qt}$, $O^8_{Qt}$, $O^8_{QQ}$ were obtained for the first time. Based on the expressions of the obtained partial amplitudes, graphs of the perturbative unitarity boundary for the listed operators were drawn. Such graphs were drawn for various operating modes of the LHC (13 TeV, 27 TeV, 100 TeV). The analysis of the impact of restrictions, following from the unitarity, on the accuracy of extracted Wilson coefficients was conducted.

Constraints, obtained from combined statistics of both channels, were also presented. The results in summary Table~\ref{Tab:summary} have been calculated by taking into account EW contribution in three top quark production at LO, QCD NLO contribution in four top quark production, and with applied partial unitarity requirements discussed in the text above. The expected limits correspond to the univariate variation of each Wilson coefficient separately. 
Results show that in the case of the operator $O_{tt}^1$, better constraints on the corresponding Wilson coefficient are obtained from the process of four top quark production. For other four operators $O_{QQ}^1$, $O_{Qt}^1$, $O_{Qt}^8$ and $O_{QQ}^8$, both processes have similar sensitivity. Combined statistics of three and four top quark production processes provide the most precise constraints in all considered cases.
The sensitivity to triple top quark production in present collider experiments is largely limited by statistical uncertainty. For future colliders, such as HL-LHC, the statistical uncertainty will be comparable to or much smaller than the systematic uncertainty, and the sensitivity to the BSM contribution in triple top quark production will be limited by theoretical calculations and experimental uncertainties. In Table~\ref{Tab:summary} a conservative estimation of the uncertainties is used for future colliders, and the partial unitarity bounds also reduce the sensitivity for the considered EFT operators. As a result, we do not observe a dramatic increase in the sensitivity of three and four top quark production processes to EFT operators with significant increases in the energy of future colliders.

Overall, the production of three top quarks seems to be quite an interesting target for BSM studies. Despite the lower cross-section (as compared to the more widely discussed four top quark production), it can potentially provide some interesting opportunities for constraining the New Physics.
Experimental challenges to distinguish between triple top-quark and four top-quark production processes can be partially overcome by splitting the phase space using a kinematic neural network and simulating a complete set of diagrams for the $p,p\to t,\bar{t},t,W,\bar{b}$ process, similar to what was done for the tWb process~\cite{boos2023separation631908339}.

\section*{Acknowledgments}

This work was supported by the Russian Science Foundation [grant number 22-12-00152].

\section*{Appendix A}
Leading helicity and partial amplitudes for process $tt\to tt$:\\

operator case $O_{tt}^{1}$:

\begin{align}\label{unit_formula5}
	A = \left(\frac{C_{tt}^{1}}{\Lambda^2}\right)\cdot 2\cdot s\cdot(1+\beta)^2
\end{align}

\begin{align}\label{unit_formula6}
	a_0 = \frac{1}{16\pi\cdot s\cdot\beta}\left| \int\limits^{0}_{4M_t^2-s}dt\cdot A \right | =  \left(\frac{C_{tt}^{1}}{\Lambda^2}\right)\cdot\frac{s}{8\pi}\cdot \beta(1+\beta)^2
\end{align}
where:
\begin{align}\label{unit_formula7}
	\beta = \sqrt{1-\frac{4M_t^2}{s}}
\end{align}

operator case $O_{QQ}^{1}$:

\begin{align}\label{unit_formula8}
	A = \left(\frac{C_{QQ}^{1}}{\Lambda^2}\right)\cdot 2\cdot s\cdot(1+\beta)^2
\end{align}

\begin{align}\label{unit_formula9}
	a_0 =  \left(\frac{C_{QQ}^{1}}{\Lambda^2}\right)\cdot\frac{s}{8\pi}\cdot \beta(1+\beta)^2
\end{align}

operator case $O_{Qt}^{1}$:

\begin{align}\label{unit_formula10}
	A = \left(\frac{C_{Qt}^{1}}{\Lambda^2}\right)\cdot 2\cdot s\cdot(1+\beta^2)
\end{align}

\begin{align}\label{unit_formula11}
	a_0 =  \left(\frac{C_{Qt}^{1}}{\Lambda^2}\right)\cdot\frac{s}{8\pi}\cdot \beta(1+\beta^2)
\end{align}

operator case $O_{Qt}^{8}$:

\begin{align}\label{unit_formula12}
	A = \left(\frac{C_{Qt}^{8}}{\Lambda^2}\right)\cdot \frac{\lambda^a_{j,i}\lambda^a_{k,l}}{4}\cdot 2\cdot s\cdot(1+\beta^2) = \left(\frac{C_{Qt}^{8}}{\Lambda^2}\right)\cdot s\cdot(1+\beta^2)
\end{align}

\begin{align}\label{unit_formula13}
	a_0 =  \left(\frac{C_{Qt}^{8}}{\Lambda^2}\right)\cdot\frac{s}{16\pi}\cdot \beta(1+\beta^2)
\end{align}

operator case $O_{QQ}^{8}$:

\begin{align}\label{unit_formula14}
	A = \left(\frac{C_{QQ}^{8}}{\Lambda^2}\right)\cdot \frac{\lambda^a_{j,i}\lambda^a_{k,l}}{4}\cdot 2\cdot s\cdot(1+\beta)^2 = \left(\frac{C_{QQ}^{8}}{\Lambda^2}\right)\cdot s\cdot(1+\beta)^2
\end{align}

\begin{align}\label{unit_formula15}
	a_0 =  \left(\frac{C_{QQ}^{8}}{\Lambda^2}\right)\cdot\frac{s}{16\pi}\cdot \beta(1+\beta)^2
\end{align}
\\

Leading helicity and partial amplitudes for process $t\bar{t}\to t\bar{t}$:\\

operator case $O_{tt}^{1}$:

\begin{align}\label{unit_formula16}
	A = \left(\frac{C_{tt}^{1}}{\Lambda^2}\right)\cdot 2\cdot u\cdot\frac{(1+\beta)}{\beta}
\end{align}

\begin{align}\label{unit_formula17}
	a_0 = \left(\frac{C_{tt}^{1}}{\Lambda^2}\right)\cdot\frac{s}{16\pi}\cdot \beta^2(1+\beta)
\end{align}

operator case $O_{QQ}^{1}$:

\begin{align}\label{unit_formula18}
	A = \left(\frac{C_{tt}^{1}}{\Lambda^2}\right)\cdot 2\cdot u\cdot\frac{(1+\beta)}{\beta}
\end{align}

\begin{align}\label{unit_formula19}
	a_0 =  \left(\frac{C_{QQ}^{1}}{\Lambda^2}\right)\cdot\frac{s}{16\pi}\cdot \beta^2(1+\beta)
\end{align}

operator case $O_{Qt}^{1}$:

\begin{align}\label{unit_formula20}
	A = \left(\frac{C_{Qt}^{1}}{\Lambda^2}\right)\cdot \left(2\cdot u\cdot\frac{(1-\beta^2)}{\beta^2} - 2\cdot s\cdot(1+\beta^2)\right)
\end{align}

\begin{align}\label{unit_formula21}
	a_0 =  \left(\frac{C_{Qt}^{1}}{\Lambda^2}\right)\cdot\left(\frac{s}{8\pi}\cdot \beta(1+\beta^2) - \frac{s}{16\pi}\cdot \beta(1-\beta^2)\right)
\end{align}

operator case $O_{Qt}^{8}$:

\begin{align}\label{unit_formula22}
	A = \left(\frac{C_{Qt}^{1}}{\Lambda^2}\right)\cdot \frac{\lambda^a_{j,i}\lambda^a_{k,l}}{4}\cdot \left(2\cdot u\cdot\frac{(1-\beta^2)}{\beta^2} - 2\cdot s\cdot(1+\beta^2)\right)
\end{align}

\begin{align}\label{unit_formula23}
	a_0 =  \left(\frac{C_{Qt}^{1}}{\Lambda^2}\right)\cdot\left(\frac{s}{16\pi}\cdot \beta(1+\beta^2) - \frac{s}{32\pi}\cdot \beta(1-\beta^2)\right)
\end{align}

operator case $O_{QQ}^{8}$:

\begin{align}\label{unit_formula24}
	A = \left(\frac{C_{tt}^{1}}{\Lambda^2}\right)\cdot \frac{\lambda^a_{j,i}\lambda^a_{k,l}}{4}\cdot 2\cdot u\cdot\frac{(1+\beta)}{\beta}
\end{align}

\begin{align}\label{unit_formula25}
	a_0 =  \left(\frac{C_{QQ}^{8}}{\Lambda^2}\right)\cdot\frac{s}{32\pi}\cdot \beta^2(1+\beta)
\end{align}


\bibliographystyle{unsrt}
\bibliography{34t_eft_v3.bib}

\end{document}